\documentclass[12pt]{article}
\usepackage[sectionbib]{natbib}
\usepackage{array,epsfig,rotating}
\usepackage{hyperref}  
\usepackage{bm}
\usepackage[ruled,linesnumbered]{algorithm2e}
\usepackage{subcaption}
\usepackage{graphicx,color}
\usepackage{booktabs}
\usepackage{diagbox}
\usepackage{multirow}
\usepackage{sectsty, secdot}
\sectionfont{\fontsize{12}{14pt plus.8pt minus .6pt}\selectfont}
\renewcommand{\theequation}{\thesection\arabic{equation}}
\subsectionfont{\fontsize{12}{14pt plus.8pt minus .6pt}\selectfont}
\usepackage{xcolor}

\usepackage{amsmath}
\usepackage{amssymb}
\usepackage{amsfonts}
\usepackage{multirow}
\usepackage{amsthm}

\newtheorem{theorem}{Theorem}
\newtheorem{lemma}{Lemma}
\newtheorem{corollary}{Corollary}

\theoremstyle{definition}
\newtheorem{definition}{Definition}
\newtheorem{assumption}{Assumption}

\usepackage{fancyhdr}

\begin{document}

\renewcommand{\baselinestretch}{2}

\markright{ \hbox{\footnotesize\rm Statistica Sinica
}\hfill\\[-13pt]
\hbox{\footnotesize\rm
}\hfill }

\renewcommand{\thefootnote}{}
$\ $\par


\fontsize{12}{14pt plus.8pt minus .6pt}\selectfont \vspace{0.8pc}
\centerline{\large\bf CONSISTENT COMMUNITY DETECTION IN}
\vspace{2pt} 
\centerline{\large\bf MULTI-LAYER NETWORKS WITH HETEROGENEOUS}
\vspace{2pt} 
\centerline{\large\bf DIFFERENTIAL PRIVACY}
\vspace{.4cm} 
\centerline{Yaoming Zhen$^a$, Shirong Xu$^b$, and Junhui Wang$^c$} 
\vspace{.4cm} 
\centerline{\it Department of Statistics, The Chinese University of Hong Kong$^{a,c}$}
\centerline{\it Department of Statistics and Data Science, University of California, Los Angeles$^b$}
\vspace{.55cm} \fontsize{9}{11.5pt plus.8pt minus.6pt}\selectfont

\fancyhf{}
\renewcommand{\baselinestretch}{2}

 \def\n{\noindent}
\chead[]{}
\fancyfoot[C]{\thepage}
\newcommand\shorttitle{MULTI-LAYER NETWORKS WITH DIFFERENTIAL PRIVACY}
\newcommand\authors{Y. ZHEN, S. XU, AND J. WANG}
\fancyhead[C]{%
\ifodd\value{page}
  \small\scshape\shorttitle
\else
  \small\scshape\authors
\fi }
\markright{ \hbox{\footnotesize\rm Statistica Sinica
}\hfill\\[-13pt]
\hbox{\footnotesize\rm
}\hfill }



\begin{quotation}
\noindent {\it Abstract:}
  As network data has become increasingly prevalent, a substantial amount of attention has been paid to the privacy issue in publishing network data. One of the critical challenges for data publishers is to preserve the topological structures of the original network while protecting sensitive information. In this paper, we propose a personalized edge flipping mechanism that allows data publishers to protect edge information based on each node's privacy preference. It can achieve differential privacy while preserving the community structure under the multi-layer degree-corrected stochastic block model after appropriately debiasing, and thus consistent community detection in the privatized multi-layer networks is achievable. Theoretically, we establish the consistency of community detection in the privatized multi-layer network and show that better privacy protection of edges can be obtained for a proportion of nodes while allowing other nodes to give up their privacy. Furthermore, the advantage of the proposed personalized edge-flipping mechanism is also supported by its numerical performance on various synthetic networks and a real-life multi-layer network.

\vspace{9pt}
\par
\end{quotation}\par

\def\thefigure{\arabic{figure}}
\def\thetable{\arabic{table}}

\renewcommand{\theequation}{\thesection.\arabic{equation}}

\fontsize{12}{14pt plus.8pt minus .6pt}\selectfont

\section{Introduction}
Network data has arisen as one of the most popular data formats in the past decades, providing an efficient way to represent complex systems involving various entities and their pairwise interactions. Among its wide spectrum of applications, the most notable examples reside in social networks \citep{du2007community, leskovec2010empirical,abawajy2016privacy}, which have been frequently collected by social network sites including Facebook, Twitter, LinkedIn, and Sina Weibo, and then published to third party consumers for academic research \citep{granovetter2005impact,li2013applications}, advertisement \citep{klerks2004network,gregurec2011importance}, crime analysis \citep{carrington2011crime, ji2014structure}, and other possible purposes. However, social network data usually conveys sensitive information related to users' privacy, and releasing them to public will inevitably lead to privacy breach, which may be abused for spam or fraudulent behaviors \citep{thomas2010koobface}. Therefore, it is imperative to obfuscate network data to avoid privacy breach without compromising the intrinsic topological structures of the network data.

To protect privacy of data, differential privacy has emerged as a standard framework for measuring the capacity of a randomized algorithm in terms of privacy protection. Its applications to network data are mainly concentrated on two scenarios, node differential privacy \citep{kasiviswanathan2013analyzing,day2016publishing,ullman2019efficiently} and edge differential privacy \citep{karwa2016inference,hehir2022consistent,yan2021directed, yan2023differentially}. The former aims to protect the privacy of all edges of some nodes while the latter mainly focuses on limiting the disclosure of edges in networks. A critical challenge in privacy-preserving network data analysis lies in understanding the effect of privacy guarantee on the subsequent data analyses, such as community detection \citep{hehir2022consistent}, degree inference \citep{yan2021directed}, and link prediction \citep{xu2018dpne,epasto2022differentially}.

In this paper, we investigate a scenario where a multi-layer network is shared with third parties for community detection while preserving edge privacy. Although numerous methods have been proposed for community detection in multi-layer networks \citep{lei2020consistent,chen2022global,xu2023covariate,ma2023community}, the privacy implications in this context remain largely unexplored in the literature. Moreover, existing network data analyses predominantly consider providing uniform privacy protection for edges within single-layer networks, disregarding the heterogeneous privacy preferences of users in practical scenarios. These approaches not only diminish the service quality for users willing to fully give up their privacy but also offer inadequate protection for those who are more concerned about their privacy. To address this challenge, we introduce a personalized edge-flipping mechanism designed to accommodate the diverse privacy preferences of individual users. It empowers users to specify the level of connectivity behavior they are comfortable sharing within a social network. Thus, our approach enables the release of networks with varying degrees of privacy protection on edges. Notably, we find that the community structure of the privatized network remains consistent through appropriate debiasing procedure under the degree-corrected multi-layer stochastic block model (DC-MSBM), preserving the utility of the original network for community detection. Correspondingly, we develop a community detection method tailored for privatized multi-layer networks and establish its theoretical guarantees for community detection consistency. Our theoretical findings are reinforced through experimentation on synthetic networks and the FriendFeed network.

The rest of the paper is structured as follows. Section \ref{Sec:pre} introduces the notations of tensors and the background of DC-MSBM. Section \ref{Sec:DPNet} introduces the application of differential privacy in network data. In Section \ref{Sec:Method}, we propose the personalized edge-flipping mechanism and show that the community structure of DC-MSBM stays invariant under this mechanism, for which we develop an algorithm for community detection on privatized networks. Section \ref{Sec:Theory} establishes the consistency of community detection of the proposed method. Section \ref{Sec:Simu} conducts various simulations to validate the theoretical results and apply the proposed method to a FriendFeed network. Section \ref{Sec:Conc} concludes the paper, and all technical proofs and necessary lemmas are deferred to the Appendix.

\section{Preliminaries}
\label{Sec:pre}

\subsection{Background of Multi-layer Networks}
We first introduce some notations on tensors, as well as some basics of DC-MSBM; \citealt{paul2021null}). Throughout the paper, we denote $[n] = \{1, ..., n\}$ for any positive integer $n$,  and denote tensors by bold Euler script letters. For a tensor $\bm{\mathcal{A} }\in \mathbb{R}^{I_1 \times I_2 \times I_3}$, denote $\bm{\mathcal{A}}_{i_1,:,:} \in \mathbb{R}^{I_2 \times I_3}$, $\bm{\mathcal{A}}_{:,i_2,:} \in \mathbb{R}^{I_1 \times I_3}$ and $\bm{\mathcal{A} }_{:,:,i_3} \in \mathbb{R}^{I_1 \times I_2}$ as the $i_1$-th horizontal, $i_2$-th lateral, and $i_3$-th frontal slide of $\bm{\mathcal{A}}$, respectively. In addition, denote $\bm{\mathcal{A}}_{:,i_2,i_3} \in \mathbb{R}^{I_1}$, $\bm{\mathcal{A}}_{i_1,:,i_3} \in \mathbb{R}^{I_2}$, and $\bm{\mathcal{A}}_{i_1,i_2,:} \in \mathbb{R}^{I_3}$ as the $(i_2, i_3)$-th mode-$1$, $(i_1, i_3)$-th mode-$2$ and $(i_1, i_2)$-th mode-$3$ fiber of $\bm{\mathcal{A}}$, respectively. For $j\in [3]$, let $\bm{\mathcal{M}}_j (\bm{\mathcal{A}})$ be the mode-$j$ major matricization of $\bm{\mathcal{A}}$ \citep{kolda2009tensor}. Specifically, $\bm{\mathcal{M}}_j(\bm{\mathcal{A})} $ is a matrix in $\mathbb{R}^{I_j \times \prod_{i \neq j}I_i}$ such that 
$$\bm{\mathcal{A}}_{i_1,i_2,i_3} = \big( \bm{\mathcal{M}}_j(\bm{\mathcal{A}})\big)_{i_j, m}, \text{ with } m = 1 + \sum_{\substack{l= 1\\ l\ne j}}^3(i_l-1)\prod_{\substack{i= 1\\ i\neq j}}^{l-1}I_i.$$ For some matrices $\bm{M}^{(1)} \in \mathbb{R}^{J_1\times I_1}$, $\bm{M}^{(2)} \in \mathbb{R}^{J_2\times I_2}$, $\bm{M}^{(3)} \in \mathbb{R}^{J_3\times I_3}$, the mode-$1$ product between $\bm{\mathcal{A}}$ and $\bm{M}^{(1)}$ is a $J_1 \times I_2 \times I_3$ tensor, defined as $(\bm{\mathcal{A}} \times_1 \bm{M}^{(1)})_{j_1, i_2, i_3} = \sum_{i_1 =1}^{I_1} \bm{\mathcal{A}}_{i_1,i_2,i_3}\bm{M}^{(1)}_{j_1, i_1}$, for $j_1 \in [J_1]$, $i_2 \in [I_2]$, and $i_3\in [I_3]$. The mode-$2$ product $\bm{\mathcal{A}} \times_2 \bm{M}^{(2)} \in \mathbb{R}^{I_1\times J_2 \times I_3}$ and mode-$3$ product $\bm{\mathcal{A}} \times_3 \bm{M}^{(3)} \in \mathbb{R}^{I_1 \times I_2 \times J_3}$ are defined similarly. The Tucker rank, also known as multi-linear rank, of $\bm{\mathcal{A}}$ is defined as $(r_1, r_2, r_3)$, where $r_1 = \text{rank}(\bm{\mathcal{M}}_1(\bm{\mathcal{A}}))$, $r_2 = \text{rank}(\bm{\mathcal{M}}_2(\bm{\mathcal{A}}))$ and $r_3 = \text{rank}(\bm{\mathcal{M}}_3(\bm{\mathcal{A}}))$. Further, if $\bm{\mathcal{A}}$ has Tucker rank $(r_1, r_2, r_3)$, it admits the following Tucker decomposition,
$$
\bm{\mathcal{A}} = \bm{\mathcal{C}} \times_1 \bm{U} \times_2 \bm{V} \times_3 \bm{W},
$$
where $\bm{\mathcal{C}} \in \mathbb{R}^{r_1 \times r_2 \times r_3}$ is a core tensor and $\bm{U} \in \mathbb{R}^{I_1 \times r_1}$,
$\bm{V} \in \mathbb{R}^{I_2 \times r_2}$ and $\bm{W} \in \mathbb{R}^{I_3 \times r_3}$ have orthonormal columns.

Let $\mathcal{G} = (\mathcal{V},\mathcal{E})$ denote a multi-layer network with $\mathcal{V}= [n]$ being the set of $n$ nodes and $\mathcal{E} = \{ \bm{E}^{(l)}\}_{l=1}^L$ being the edge sets for all $L$ layers, where $(i,j) \in \bm{E}^{(l)}$ if there exists an edge between nodes $i$ and $j$ in the $l$-th network layer. Generally, $\mathcal{G}$ can be equivalently represented by an order-3 adjacency tensor $\bm{\mathcal{A}} \in \{0,1 \}^{n \times n \times L}$ with $\bm{\mathcal{A}}_{:,:,l} = \bm A^{(l)}$, where $\bm{A}_{i,j}^{(l)} = \bm{A}_{j, i}^{(l)}=1$ if $(i,j) \in \bm{E}^{(l)}$ and $\bm{A}_{i,j}^{(l)} = \bm{A}_{j, i}^{(l)}=0$ otherwise. Moreover, we denote by $\bm{\mathcal{P}}$ the underlying probability tensor such that $\bm{\mathcal{P}}_{:,:,l} =\bm{P}^{(l)}$ with $\bm{P}^{(l)}_{i,j} = P(\bm{A}_{i,j}^{(l)} = 1)$ denoting the probability that there exists an edge between nodes $i$ and $j$ in the $l$-th network layer. The DC-MSBM model essentially assumes that
\begin{align*}
	\bm{P}_{i,j}^{(l)} = d_i d_j \bm{B}_{c_i,c_j}^{(l)}, \mbox{ for } i,j\in [n], l\in [L],
\end{align*}
where $c_i$ and $d_i$  denote the community membership assignment and degree heterogeneity parameter of node $i$ across all network layers, and $\bm{B}_{c_i,c_j}^{(l)}$ is the linking probability between community $c_i$ and $c_j$ in the $l$-th layer. Note that we assume the community memberships of the nodes are homogeneous across all network layers. This allows us to define a community membership matrix. Specifically, let $\bm Z \in \{0,1 \}^{n \times K}$ be the community membership matrix such that ${\bm Z}_{i, c_i}=1$ and ${\bm Z}_{i, k}=0$ for $k \ne c_i$. The probability tensor of the DC-MSBM can then be written as
\begin{align}
	\label{equ:dcbm}
	\bm{\mathcal{P}} = \bm{\mathcal{ B}} \times_1 \bm{D}\bm Z \times_2 \bm{D}\bm Z,
\end{align}
where $\bm{D} = diag\{d_1,\ldots,d_n\}$ is a diagonal matrix and $\bm{\mathcal{ B}} \in  [0,1]^{K \times K \times L}$ is the core probability tensor with $\bm{\mathcal{ B}}_{:, :, l} = \bm B^{(l)}$.

Furthermore, for two sequences $f_n$ and $g_n$, we denote $f_n = O(g_n)$ if $\lim_{n\rightarrow +\infty}\sup |f_n|/g_n < +\infty$, $f_n = o(g_n)$ if $\lim_{n\rightarrow +\infty} |f_n|/g_n =0$, $f_n = \Omega(g_n)$ if $\lim_{n\rightarrow +\infty} \sup |f_n|/g_n >0$, $f_n\gg g_n$ if $\lim_{n\rightarrow +\infty} |f_n|/g_n = +\infty$, and $f_n \asymp g_n$ if $f_n = O(g_n)$ and $f_n = \Omega(g_n)$. Let $\Vert \cdot \Vert$ denote the $l_2$-norm of a vector or the spectral norm of a matrix, $\Vert \cdot \Vert_\infty$ denote the $l_\infty$-norm of the vectorization of the input matrix or tensor, and $\Vert \cdot \Vert_F$ denote the Frobenius norm of a matrix or tensor, and the $l_{2, 1}$-norm of a matrix $\bm{M}\in \mathbb{R}^{r\times c}$ is defined as $\Vert \bm{M} \Vert_{2,1} = \sum_{i=1}^r \Vert \bm{M}_{i,:}\Vert$, where $\bm{M}_{i, :}$ is the $i$-th row of $\bm{M}$.

\subsection{Differential Privacy}
Differential privacy (DP; \citealt{dwork2006calibrating}) has emerged as a standard statistical framework for protecting personal data during data sharing processes. The formal definition of $\epsilon$-DP is given as follows.

\begin{definition}[$\epsilon$-DP]
A randomized mechanism $\mathcal{M}$ satisfies $\epsilon$-differential privacy if for any two datasets $\mathcal{D}$ and $\mathcal{D}^\prime$ differing in only one record,
    \begin{align*}
    \sup_{S \in \mathcal{S}}
        \frac{P(\mathcal{M}(\mathcal{D})=S)}{P(\mathcal{M}(\mathcal{D}')=S)}
        \leq \exp(\epsilon),
    \end{align*}
    where $\mathcal{S}$ denotes the output space of $\mathcal{M}$.
\end{definition}
Another variant of differential privacy is known as local differential privacy (LDP), wherein each individual data point undergoes perturbation with noise prior to data collection procedure. The formal definition of noninteractive $\epsilon$-LDP is provided as follows.
\begin{definition}[$\epsilon$-LDP]
For a given privacy parameter $\epsilon>0$, the randomized mechanism $\mathcal{M}$ satisfies $\epsilon$-local differential privacy for $X$ if
    	\begin{align*}
		\sup_{\widetilde{x} \in \mathcal{X}} \sup_{x,x' \in \mathcal{X}}\frac{P\big(\mathcal{M}(X)=\widetilde{x}|X=x\big)}{P\big(\mathcal{M}(X)=\widetilde{x}|X=x'\big)} \leq \exp(\epsilon),
	\end{align*}
 where $\mathcal{X}$ denotes the output space of $\mathcal{M}$.
\end{definition}

It's worth noting that privacy protection under $\epsilon$-LDP can be analyzed within the framework of classic $\epsilon$-DP in specific scenarios. Specifically, if $\mathcal{M}$ satisfies $\epsilon$-local differential privacy and is applied to samples of a dataset independently, then the following holds
\begin{align*}
\sup_{\widetilde{\mathcal{D}} \in \mathcal{X}^n}
\frac{P\big(\mathcal{M}(\mathcal{D})=\widetilde{\mathcal{D}}\big)}{P\big(\mathcal{M}(\mathcal{D}')=\widetilde{\mathcal{D}}\big)}
=
\sup_{\widetilde{x} \in \mathcal{X}}
\frac{P\big(\mathcal{M}(X_i)=\widetilde{x}|X_i = x\big)}{P\big(\mathcal{M}(X_i)=\widetilde{x}|X_i = x'\big)} \leq \exp(\epsilon),
\end{align*}
where $\mathcal{D}$ and $\mathcal{D}'$ differ in the $i$-th record. Thus, $\epsilon$-LDP achieves the classic $\epsilon$-DP if we consider the output space $\mathcal{S}=\mathcal{X}^n$.

\section{Differential Privacy in Multilayer Networks}
\label{Sec:DPNet}
In the realm of network data, two primary variants of differential privacy emerge: node differential privacy \citep{kasiviswanathan2013analyzing,day2016publishing} and edge differential privacy \citep{karwa2016inference,wang2022two,yan2023differentially}. The former considers the protection of all information associated with a node in network data, while the latter on the edges. This paper delves into the privacy protection of edges in multi-layer networks. The formal definition of $\epsilon$-edge differential privacy is given as follows.

\begin{definition}
    ($\epsilon$-edge DP)
A randomized mechanism $\mathcal{M}$ is $\epsilon$-edge differentially private if
\begin{align*}
\sup_{S \in \mathcal{S}}
\sup_{\delta(\bm{\mathcal{A}},\bm{\mathcal{A}}')=1}
    \frac{P(\mathcal{M}(\bm{\mathcal{A}})=S|\bm{\mathcal{A}})}{P(\mathcal{M}(\bm{\mathcal{A}}')=S|\bm{\mathcal{A}'})}
    \leq \exp(\epsilon),
\end{align*}
where $\bm{\mathcal{A}}$ and $\bm{\mathcal{A}}'$ are two neighboring multi-layer networks differing in one edge and $\mathcal{S}$ denotes the output space of $\mathcal{M}(\cdot)$.
\end{definition}

The definition of $\epsilon$-edge DP bears a resemblance to classic $\epsilon$-DP, as it requires the output distribution of the randomized mechanism $\mathcal{M}$ to remain robust against alterations to any single edge in the network. It is thus difficult for attackers to infer any single edge based on the released network information $S$. In the literature, $\epsilon$-edge DP finds widespread use in releasing various network information privately, such as node degrees \citep{karwa2016inference,fan2020asymptotic}, shortest path length \citep{chen2014correlated}, and community structure \citep{mohamed2022differentially}. Under the framework of $\epsilon$-LDP, we consider a specific variant of $\epsilon$-LDP for edges in multilayer network data.  
\begin{definition}
	\label{def:epsilon_edge LDP}
	($\epsilon$-edge local differential privacy) Let $\bm{\mathcal{A}}$ denote the adjacency tensor of a multi-layer network with $n$ common nodes. We say a randomized mechanism $\mathcal{M}$ satisfies $\epsilon$-edge local differential privacy if
	\begin{equation}
		\label{equ:edge_dp}
		\sup_{\widetilde{x}\in  \mathcal{X}} \sup_{x,x' \in \mathcal{X}}\frac{P\big(\mathcal{M}(\bm{\mathcal{A}}_{i,j,l})=\widetilde{x}|\bm{\mathcal{A}}_{i,j,l}=x\big)}{P\big(\mathcal{M}(\bm{\mathcal{A}}_{i,j,l})=\widetilde{x}|\bm{\mathcal{A}}_{i,j,l}=x'\big)} \leq \exp(\epsilon),
	\end{equation}
 where $\mathcal{X}$ denotes the range of edges
\end{definition}

Clearly, the local version of $\epsilon$-edge DP is intrinsically connected to its central counterpart. Particularly, if $\mathcal{M}$ satisfies $\epsilon$-edge LDP and is applied to $\bm{\mathcal{A}}$ entrywisely. 
Given the independence of $\bm{\mathcal{A}}_{i,j,l}$'s, we have
\begin{align}
\label{Eqi:local}
&\sup_{\bm{\mathcal{\widetilde{A}}} \in \mathcal{X}^{n\times n\times L}}
\sup_{\delta(\bm{\mathcal{A}} ,\bm{\mathcal{A}}^\prime)=1} \frac{P\big(\mathcal{M}(\bm{\mathcal{A}}) = \bm{\mathcal{\widetilde{A}}}|\bm{\mathcal{A}} \big)}{P\big(\mathcal{M}(\bm{\mathcal{A}}^\prime) = \bm{\mathcal{\widetilde{A}}}|\bm{\mathcal{A}}^\prime\big)}\notag \\
=&
\sup_{i,j,l}
\sup_{\widetilde{x}\in  \mathcal{X}} \sup_{x,x' \in \mathcal{X}}\frac{P\big(\mathcal{M}(\bm{\mathcal{A}}_{i,j,l})=\widetilde{x}|\bm{\mathcal{A}}_{i,j,l}=x\big)}{P\big(\mathcal{M}(\bm{\mathcal{A}}_{i,j,l})=\widetilde{x}|\bm{\mathcal{A}}_{i,j,l}=x'\big)} \leq \exp(\epsilon).
\end{align}

It is evident from (\ref{Eqi:local}) that privacy protection through $\epsilon$-edge LDP is equivalent to achieving $\epsilon$-edge DP, provided the independence of edges. In simpler terms, a randomized mechanism $\mathcal{M}$ satisfying $\epsilon$-edge LDP can be regarded as a specialized method for achieving $\epsilon$-edge DP, wherein the output is a new multi-layer network. Furthermore, a similar correlation can be established between $\epsilon$-edge LDP and the $(k, \epsilon)$-edge DP framework, as explored in prior works such as \cite{hay2009accurate} and \cite{yan2023differentially}.

In this paper, we mainly consider multilayer networks with binary edges, i.e., $\mathcal{X}=\{0,1\}$. To achieve $\epsilon$-edge LDP, one popular choice of $\mathcal{M}$ is the edge-flipping mechanism of $\bm{\mathcal{A}}$ with a uniform flipping probability \citep{nayak2009unified,wang2016using,hehir2022consistent}. Specifically, denote the flipped multi-layer network as $\mathcal{M}_{\theta}(\bm{\mathcal{A}})$ with a flipping probability $1-\theta$, for some $\theta\ge 1/2$, then the $(i, j, l)$-th entry of $\mathcal{M}(\bm{\mathcal{A}})$ is given by
\begin{align*}
	\mathcal{M}_{\theta}(\bm{\mathcal{A}}_{i,j,l}) = 
	\begin{cases}
		\bm{\mathcal{A}}_{i,j,l}, &\mbox{ with probability } \theta, \\
		1-\bm{\mathcal{A}}_{i,j,l},& \mbox{ with probability }  1-\theta.
	\end{cases}
\end{align*}
It then follows that $P\big(\mathcal{M}_{\theta}(\bm{\mathcal{A}})_{i,j,l}=1\big) = \theta \bm{\mathcal{P}}_{i,j,l} + (1-\theta)(1-\bm{\mathcal{P}}_{i,j,l})$. 


\begin{lemma}
	\label{FlipMec}
	The edge-flipping mechanism $\mathcal{M}_{\theta}$ satisfies $\epsilon$-edge local differential privacy when $\theta =  \frac{e^{\epsilon}}{1+e^{\epsilon}}$.
\end{lemma}

Lemma $\ref{FlipMec}$ characterizes the capacity of the edge-flipping mechanism in protecting privacy under the framework of $\epsilon$-edge LDP. It should be noted that privacy of $\bm{\mathcal{A}}$ is completely protected when $\theta=1/2$ or $\epsilon = 0$, in the sense that there exists no algorithm capable of inferring $\bm{\mathcal{A}}_{i,j,l}$ based on $\mathcal{M}_{1/2}(\bm{\mathcal{A}}_{i,j,l})$ more effectively than random guessing. Yet, a key disadvantage of the uniform flipping mechanism is its inability to accommodate different privacy preferences among edges.

We further emphasize that the $\epsilon$-edge LDP achieved by the edge-flipping mechanism $\mathcal{M}$ remains consistent with the definition of $\epsilon$-edge DP. The data we release is the privacy-preserving network after random flipping, presented as a unified tensor, despite its composition of $O(n^2L)$ edges. In contrast to mechanisms that solely disclose summary statistics of the network, our approach enables the release of a complete data tensor with the same expected expectation as the original network after a debiasing step. However, it is important to note that some structures of the original network cannot be recovered directly due to privacy protection. Instead, estimations, such as the precise count of triangles in the original network, are still obtainable. In essence, the publication of the privacy-preserving network allows for releasing more data about networks.

\section{Proposed Method}
\label{Sec:Method}
\subsection{Personalized Edge-flipping}

In this section, we propose a personalized edge-flipping mechanism whose flipping probabilities are governed by node-wise privacy preferences. Specifically, let $\Theta = (\theta_{i, j})_{n \times n}$ with $\theta_{i, j}$ denoting the flipping probability of the potential edge between nodes $i$ and $j$ across all network layers, and $\mathcal{M}_{\Theta}(\bm{\mathcal{A}}) = \big(\mathcal{M}_{\theta_{i,j}}(\bm{\mathcal{A}}_{i,j,l})\big)_{n \times n\times L}$ with
\begin{align}
	\mathcal{M}_{\theta_{i,j}}(\bm{\mathcal{A}}_{i,j,l}) =
	\begin{cases}
		\bm{\mathcal{A}}_{i,j,l},  & \mbox{ with probability } \theta_{i, j}, \\
		1-\bm{\mathcal{A}}_{i,j,l}, &\mbox{ with probability } 1-\theta_{i, j},
	\end{cases}
\end{align}
for $i \leq j$ and $l \in [L]$. Also, we set $\mathcal{M}_{\theta_{i, j}}(\bm{\mathcal{A}}_{i,j,l}) = \mathcal{M}_{\theta_{j, i}}(\bm{\mathcal{A}}_{i,j,l})$ to preserve the semi-symmetry in $\mathcal{M}_{\Theta}(\bm{\mathcal{A}})$ with respect to the first two modes, for $i > j$.

\begin{definition}
	\label{def:personalized}
	(Heterogenous $\bm{\epsilon}$-edge LDP) Let $\mathcal{M}$ denote a randomized mechanism, then we say $\mathcal{M}$ satisfies heterogenous $\bm{\epsilon}$-edge LDP if for any $1\leq i<j \leq n$ and $l \in [n]$ with $\bm{\epsilon} = (\epsilon_{i,j})_{i,j=1}^n$, we have
	\begin{align*}
		\sup_{\widetilde{x}\in  \mathcal{X}} \sup_{x,x' \in \mathcal{X}}\frac{P(\mathcal{M}(\bm{\mathcal{A}}_{i,j,l})=\widetilde{x}|\bm{\mathcal{A}}_{i,j,l}=x)}{P(\mathcal{M}(\bm{\mathcal{A}}_{i,j,l})=\widetilde{x}|\bm{\mathcal{A}}_{i,j,l}=x')} \leq \exp(\epsilon_{i,j}),
	\end{align*}
	where $\epsilon_{i,j}$ is a privacy parameter depending on nodes $i$ and $j$.
\end{definition}

Compared with $\epsilon$-edge LDP, the proposed heterogenous $\bm{\epsilon}$-edge LDP allows for the variation of the privacy parameter $\epsilon_{i, j}$ from edge to edge. Particularly, heterogenous $\bm{\epsilon}$-edge LDP is equivalent to $\epsilon$-edge LDP when $\epsilon = \max_{i, j}\epsilon_{i,j}$. The developed concept bears resemblance to heterogeneous differential privacy \citep{alaggan2015heterogeneous} in nature, wherein individual points in a dataset are provided different privacy guarantees. The motivation behind heterogenous $\bm{\epsilon}$-edge LDP is to cater to the diverse preferences among users in the network. While some users may prioritize better service over privacy, others may prioritize keeping their social interactions as private as possible.

To allow for node-specified privacy preferences, we propose to parametrize $\Theta$ as 
\begin{align}
	\label{Facto_mech}
	\Theta = \frac{1}{2}(\bm{f} \bm{f}^\top + \bm{1}_n \bm{1}_n^T),
\end{align}
where $\bm{f} = (f_1, ..., f_n)^\top \in [0,1]^n$ is a vector consisting of the privacy preferences of all nodes and $\bm{1}_n$ is the vector with $n$ ones. In particular, when $f_i = 0$, it signifies that $\theta_{i,j}=1/2$ for any $j \in [n]$, indicating that the edges associated with node $i$ are protected at the utmost secrecy level. Conversely, if $f_i = f_j = 1$, it indicates that both nodes $i$ and $j$ give up their privacy, resulting in complete exposure of $\bm{\mathcal{A}}_{i,j,l}$'s to the service provider. Essentially, the privacy level of an edge between two nodes is solely determined by their respective privacy preferences, and and the edge is exposed with a higher probability when both nodes choose weaker privacy protection.  

\begin{lemma}
	\label{lemma:private_preferences}
	The personalized edge-flipping mechanism $\mathcal{M}_{\Theta}(\bm{\mathcal{A}})$ with $\Theta$ being parametrized as in $(\ref{Facto_mech})$ satisfies heterogenous $\bm{\epsilon}$-edge LDP with $\epsilon_{i,j} = \log \frac{1+f_jf_j}{1-f_if_j}$, for $i, j \in [n]$. Moreover, 
 \begin{align*}
f_i  = \sqrt{\frac{(1 - \frac{2}{1+e^{\epsilon_{i, i^\prime}}})(1 - \frac{2}{1+e^{\epsilon_{i, j}}})}{1 - \frac{2}{1+e^{\epsilon_{i^\prime, j}}}}}, 
\end{align*}
for any $i^\prime \ne j$, $i^\prime \ne i$, and $j\ne i$.
\end{lemma}

Lemma \ref{lemma:private_preferences} shows that, under the personalized edge-flipping mechanism, the privacy guarantee of any single edge is completely determined by the pair of nodes forming that particular edge. Furthermore, it is important to note that the privacy protection provided to edges via $\mathcal{M}_{\Theta}$ is contingent upon the parameterization specified in (\ref{Facto_mech}). In other words, the level of privacy protection on edges will vary with the parameterization of $\Theta$.

\subsection{Decomposition after Debiasing}

A critical challenge in releasing network data is to preserve network structure of interest while protecting privacy of edges. It is interesting to remark that the community structure is still encoded in the flipped network under personalized edge-flipping mechanism, which allows for consistent community detection on the flipped network with some appropriate debiasing procedures.

\begin{lemma}
	\label{DCBM}
	Assume that $\bm{\mathcal{A}}$ is generated from the DC-MSBM in (\ref{equ:dcbm}) and that the personalized flipping probability matrix satisfies the factorization property in (\ref{Facto_mech}), we have
	\begin{align}
     \label{equ: A_tilde]}
		\mathbb{E}\Big(\bm{\mathcal{\widetilde{A}}}_{i,j,l} \Big)
		=
		f_i f_j
		d_{i} d_j \bm{B}_{c_i, c_j }^{(l)}, i,j \in [n], l\in [L],
	\end{align}
	where $\bm{\mathcal{\widetilde{A}}}_{i,j,l} =\mathcal{M}_{\theta_{i, j}}(\bm{\mathcal{A}}_{i,j,l}) + \frac{1}{2}(f_if_j - 1)$.
\end{lemma}

Lemma $\ref{DCBM}$ shows that the expectation of the flipped network $\mathcal{M}_{\bm{\Theta}}(\bm{\mathcal{A}})$ preserves the same community structure in $\bm{\mathcal{A}}$ after debiasing, suggesting that consistent community detection shall be conducted on the debiased network $\bm{\mathcal{\widetilde{A}}}$. 

We remark that various network data analysis tasks remain feasible after an additional debiasing step, including estimating the counts of specific sub-graphs such as $k$-stars or triangles, as well as inferring the degree sequence. This is achievable because we can obtain a tensor sharing exactly the same expectation as $\bm{\mathcal{A}}$ by further dividing $\bm{\mathcal{\widetilde{A}}}_{i, j, l}$ by $f_i f_j$, for any $i, j \in [n]$, and $l \in [L]$. For community detection, this step is not necessary, as $f_i d_i$ can be considered a new degree heterogeneous parameter for node $i$, which will be normalized in the tensor-based variation of the SCORE method \cite{jin2015fast, ke2019community}. Estimating and inferring certain network statistics on the differentially private network after debiasing is commonly employed. For example, randomized algorithms in \citet{hay2009accurate, karwa2016inference, yan2021directed, yan2023differentially} release a perturbed degree sequence, or two perturbed bi-degree sequences, or degree partitions if the order of nodes is not crucial in downstream analysis, by adding discrete Laplacian noise. Subsequently, the parameters in the $\beta$-model, with or without covariates, can be estimated using the denoised degree sequences.

It follows from Lemma \ref{DCBM} that the expectation of $\bm{\mathcal{\widetilde{A}}} $ can be decomposed as
\begin{align*}
	\mathbb{E}\big(
	\bm{\mathcal{\widetilde{A}}}
	\big) = \bm{\mathcal{ B}} \times_1 \bm{F}\bm{D} \bm Z \times_2 \bm{F} \bm{D}\bm Z,
\end{align*}
where $\bm{F} = diag(\bm{f})$. For ease of notation, we denote $\bm{\mathcal{\widetilde{P}}} = \mathbb{E}\big(\bm{\mathcal{\widetilde{A}}}\big)$, and then
\begin{align}
	\label{P_rep}
	\bm{\mathcal{\widetilde{P}}} =( \bm{\mathcal{ B}}\times_1 \bm{\Gamma}  \times_2 \bm{\Gamma} )  \times_1 \bm{F} \bm{D} \bm Z\bm{\Gamma} ^{-1} \times_2 \bm{F}\bm{D}\bm Z \bm{\Gamma}^{-1},
\end{align}
where $\bm{\Gamma} = diag(\sqrt{\gamma_1},\ldots,\sqrt{\gamma_K})$ and $\gamma_k = \sum_{i=1}^n \bm{Z}_{i,k} (f_id_i)^2$ is the effective size of the $k$-th community depending on the nodes' degree heterogeneity coefficients and heterogenous privacy preference parameters. Suppose the Tucker rank of $\bm{\mathcal{ B}}\times_1 \bm{\Gamma}  \times_2 \bm{\Gamma}$ is $(K, K, L_0)$, and thus $\bm{\mathcal{ B}}\times_1 \bm{\Gamma}  \times_2 \bm{\Gamma}$ admits the following Tucker decomposition
\begin{align}
	\label{B_tucker}
	\bm{\mathcal{ B}}\times_1 \bm{\Gamma}  \times_2 \bm{\Gamma} = 
	\bm{\mathcal{C}} \times_1 \bm{O} \times_2 \bm{O} \times_3 \bm{V},
\end{align}
for a core tensor $\bm{\mathcal{C}} \in \mathbb{R}^{K\times K \times L_0}$, and the factor matrices $\bm{O}\in \mathbb{R}^{K\times K}$ and $\bm{V} \in \mathbb{R}^{L \times L_0}$ whose columns are orthonormal. Note that $\bm{FDZ\Gamma}^{-1}$ also has orthonormal columns. Plugging $(\ref{B_tucker})$ into $(\ref{P_rep})$ yields the Tucker decomposition of $\bm{\mathcal{\widetilde{P}}}$ as
\begin{align*}
	\label{P_tucker}
	\bm{\mathcal{\widetilde{P}}} =\bm{\mathcal{C}}  
	\times_1 \bm{F}\bm{D} \bm Z\bm{\Gamma} ^{-1}\bm{O}  \times_2 \bm{F}\bm{D}\bm Z \bm{\Gamma}^{-1}\bm{O} \times_3 \bm{V}.
\end{align*} 
Denote $\bm{U} =\bm{F} \bm{D} \bm Z\bm{\Gamma} ^{-1}\bm{O}$ as the mode-1 and mode-2 factor matrix in the Tucker decomposition of $\bm{\mathcal{\widetilde{P}}}$. It can be verified that $\bm{U}$ is a column orthogonal matrix with $\bm{U}^T \bm{U} = \bm{I}_K$. 

\begin{lemma}
	\label{Factor_norm}
	For any node pair $(i, j)\in [n] \times [n]$, we have $\bm{U}_{i, :}/\Vert \bm{U}_{i, :} \Vert = \bm{U}_{j, :}/\Vert \bm{U}_{j, :} \Vert$ if $c_i^* = c_j^*$ and $\big\Vert \bm{U}_{i, :}/\Vert \bm{U}_{i, :}\Vert - \bm{U}_{j, :}/\Vert \bm{U}_{j, :}\Vert\big\Vert = \sqrt{2}$ otherwise.
\end{lemma}

Lemma \ref{Factor_norm} shows that the spectral embeddings of nodes within the same communities are the same after row-wise normalization. This motivates us to propose the following Algorithm 1 to estimate  the community structure based on the Tucker decomposition of $\bm{\mathcal{\widetilde{A}}}$.


\begin{algorithm}[!htp]
	\SetKwInOut{Input}{Input}
	\SetKwInOut{Output}{Output}
	\caption{Community detection in flipped network} \label{algorithm1}
	\Input{Flipped adjacency tensor $\mathcal{M}_{\Theta}(\bm{\mathcal{A}})$,
		 privacy parameter $\bm{f}$,
		number of communities $K$,
		tolerance $\tau$\\}
	\Output{Privacy-preserving community memberships $\widehat{\bm{Z}}$\\}
	Let $
	\bm{\mathcal{\widetilde{A}}} = \mathcal{M}_{\Theta}(\bm{\mathcal{A}})+ \frac{1}{2}(
	\bm{f} \circ\bm{f} - \bm{1}_n \circ \bm{1}_n) \circ \bm{1}_L$;  \\
	Implement Tucker decomposition on $\bm{\mathcal{\widetilde{A}}}$ with Tucker rank $(K,K,L_0 = \min\{K(K+1)/2,L \})$ as $
	\bm{\mathcal{\widetilde{A}}} \approx \widehat{\bm{\mathcal{C}}} \times_1 \widehat{\bm{U}}
	\times_2  \widehat{\bm{U}}\times_3\widehat{\bm{V}}.
	$\\
	Normalized the embedding matrix 
	$
	\widehat{\bm{\widetilde{U}}}_{i,:} = \widehat{\bm{U}}_{i,:}/\Vert \widehat{\bm{U}}_{i,:}\Vert$, for $i\in [n]$. \\
	Apply an  $(1+\tau)$-optimal $K$-medians algorithm to $\widehat{\bm{\widetilde{U}}}$ to obtain a solution $(\widehat{\bm{Z}},\widehat{\bm{W}})$ that satisfies,
	\begin{align*}
		\Vert \widehat{\bm{Z}}\widehat{\bm{W}} - \widehat{\bm{\widetilde{U}}}\Vert_{2,1} 
		\leq (1+\tau) \min_{\bm{Z}\in \bm{\Delta},\bm{W} \in \mathbb{R}^{K \times K}} \Vert \bm{Z}\bm{W} - \widehat{\bm{\widetilde{U}}} \Vert_{2,1},
	\end{align*}
	where $\bm{\Delta} \subset \{ 0,1 \}^{n \times K}$ is the set of membership matrices.
\end{algorithm}

In Algorithm \ref{algorithm1}, we first conduct a debiasing operation on the flipped network $\mathcal{M}_{\Theta}(\bm{\mathcal{A}})$ to obtain $\bm{\mathcal{\widetilde{A}}}$, such that the expectation of $\bm{\mathcal{\widetilde{A}}}$ admits the same DC-MSBM as in $\bm{\mathcal{A}}$. Next, a low rank Tucker approximation of $\bm{\mathcal{\widetilde{A}}}$ is implemented to estimate the spectral embedding matrix $\widehat{\bm{U}}$. Finally, a $(1 + \tau)$-optimal K-medians algorithm is applied to the normalization version of $\widehat{\bm{U}}$, which clusters the nodes into $K$ desired communities. Herein, we follow the similar treatment in \citet{lei2015consistency} to apply the approximating K-medians algorithm for the normalized nodes' embedding, which appears to be more robust against outliers than the K-means algorithms.

\section{Theory}
\label{Sec:Theory}
In this section, we establish the asymptotic consistency of community detection on the privatized multi-layer network under the proposed personalized edge-flipping mechanism. Particularly, let $\widehat{\bm{c}}=(\widehat{c}_1,\ldots,\widehat{c}_n)$ and $\bm{c}^*=(c_1^*,c_2^*,\ldots,c_n^*)$ denote the estimated community membership vector obtained from Algorithm 1 and the true community membership vector, respectively. We assess the community detection performance with minimum scaled Hamming distance between $\hat{\bm{c}}$ and $\bm{c}^*$ under permutation \citep{jin2015fast, jing2020community, zhen2021community}. Formally, it is defined as
\begin{equation}
	\label{equ:HE}
	\text{Err}(\widehat{\bm{c}},\bm{c}^*) = 
	\min_{\bm{\pi} \in S_K}\frac{1}{n}\sum_{i=1}^n I(c_i^* = \bm{\pi}(\widehat{c}_i)),
\end{equation}
where $S_K$ is the symmetric group of degree $K$ and $I(\cdot)$ is the indicator function. Clearly, the Hamming error in $(\ref{equ:HE})$ measures the minimum fraction of nodes with inconsistent community assignments between $\widehat{\bm c}$ and $\bm{c}^*$.

To establish the consistency of community detection, the following technical assumptions are made.
\begin{assumption}
	\label{assumption:community_size}
	Let $n_k$ be the cardinality of the $k$-th true community for $k\in [K]$, and denote $n_{\max} = \max_{k \in [K]} n_k$ and $n_{\min} = \min_{k \in [K]} n_k$, then $n_{\max}= O(n_{\min})$.
\end{assumption}

\begin{assumption}
	\label{assumption:fdtradeoff}
	Let $\gamma_{\max} = \max_{k \in [K]} \gamma_k$ and $\gamma_{\min} = \min_{k\in [K]} \gamma_k$. Assume that there exists an absolute constant $C_1$ such that
	$$
	\gamma_{\max} = O(\gamma_{\min}), \text{ and } f_i^2 d_i^2 \le C_1\frac{\gamma_{c^*_i}}{n_{c^*_i}}, \text{ for } i \in [n].
	$$
\end{assumption}


\begin{assumption}
	\label{assumption_sparsity}
	Suppose that $\bm{\mathcal{B}}_{i, j, l} = O(s_n)$ for $i, j \in [n]$ and $l \in [L]$, where $s_n$ is a network sparsity coefficient that may vanish with $n$ and $L$. Moreover, we require $s_n$ satisfies
 \begin{align*}
s_n \gg \frac{1}{\overline{\psi}} \sqrt{\frac{\varphi_n \log n }{nL}}, 
\end{align*}
where $\varphi_n = 1-\min_{i \in [n]}f_i+4s_n$, and $\overline{\psi} = \frac{1}{n}\sum_{i=1}^n (f_i d_i)^2$.
\end{assumption}


\begin{assumption}
	\label{assumption_signal}
	Assume that the core tensor $\bm{\mathcal{B}} $ in the DC-MSBM model satisfies that
	$$
	\sigma_{\min}\Big(\bm{\mathcal{M}}_3\big(\bm{\mathcal{B}} )\Big) = \Omega(\sqrt{L}s_n), 
	$$
	where $\sigma_{\min}(\cdot)$ denotes the smallest non-zero singular value of a matrix.
\end{assumption}


Assumption \ref{assumption:community_size} ensures all the $K$ true communities in $\cal A$ are non-degenerate as $n$ diverges \citep{lei2020consistent, zhen2021community}. Assumption \ref{assumption:fdtradeoff} imposes a homogeneity condition on the squared product of the nodes' privacy preference parameters and the degree heterogeneity coefficients. Assumption \ref{assumption_sparsity} places a sparsity coefficient on the core probability tensor $\bm{\mathcal{B}}$ to control the overall network sparsity, which is a common assumption for network modeling \citep{ghoshdastidar2017uniform, guo2020randomized, zhen2021community}. If there is no privacy protection at all; that is, $f_1 = \ldots = f_n = 1$, which corresponds to $\epsilon_{i, j} = +\infty$, we have $\varphi_n = 4s_n$, and $s_n \gg 4\left(\frac{n}{\sum_{i=1}^n d_i^2}\right)^2 \frac{\log n }{nL} = O(\frac{\log n}{nL})$. Clearly, this reduces to the optimal sparsity assumption for consistent community detection in multi-layer network data \cite{jing2020community}. However,  if $cs_n \ll  1 - \min_{\i\in [n]} f_i$ for some constant $c$, leading to $\varphi_n\gg s_n$, the proposed network sparsity assumption is stronger than the optimal one in general. Assumption \ref{assumption_signal} assumes the smallest non-zero sigular value of $\bm{\mathcal{M}}_3(s_n^{-1} \bm{\mathcal{B}})$ should scale at least at the order of $\sqrt{L}$. This is a  mild assumption and can be satisfied if the entries of $s_n^{-1} \bm{\mathcal{B}}$ are indepedently and identically generated from some zero-mean sub-Gaussian random variables \citep{rudelson2009smallest}.


\begin{theorem}
	\label{thm:consist}
	Under Assumptions \ref{assumption:community_size}-\ref{assumption_signal}, the Hamming error of $\widehat{\bm c}$ satisfies that
	\begin{align*}
		\mathrm{Err}(\widehat{\bm{c}},\bm{c}^*) =  O_p\left(
		\left( \sum_{k=1}^K v_k \right)^{1/2}
		\frac{\sqrt{\varphi_n \log n}}{\sqrt{n L}s_n \overline{\psi}}\right), 
	\end{align*}
	where $v_k = n_k^{-2}\sum_{c_i^*=k} \gamma_k/(f_id_i)^2$. Moreover, in the simplest case that all the $\epsilon_{i, j}$'s are the same, denoted as $\epsilon$, leading to $f_i^2 = 1 - \frac{2}{1 + \exp\{\epsilon\}}$, for $i\in [n]$. The Hamming error between $\bm{\hat{c}}$ and $\bm{c}^*$ can be rewritten as 
\begin{align*}
    \quad \mathrm{Err}(\bm{\hat{c}}, \bm{c}^*)  = O_p\Bigg(\sqrt{\frac{\log n}{nLs_n^2 \epsilon^2 }}\Bigg),
\end{align*}
when $\epsilon$ is sufficiently small, provided that the degree heterogeneous parameters are asymptotically of the same order. 
\end{theorem}

Theorem 1 provides a probabilistic upper bound for the community detection error under the personalized edge-flipping mechanism. For one simple scenario with $f_i = d_i = 1$ for $i \in [n]$, where there is no privacy protection and degree heterogeneity, Theorem 1 implies that $\varphi_n \asymp s_n$ and Err$(\hat{\bm{c}}, \bm{c}^*) = o(1)$ as long as $s_n \gg \frac{\log n}{nL}$ and $K = O(1)$, which matches with the optimal sparsity requirement for consistent community detection on multi-layer networks \citep{jing2020community}. However, when $\min_{i\in [n]}f_i$ deviates from 1, $\varphi_n$ will become substantially larger than $s_n$, leading to deterioration of the convergence rate of the Hamming error. 

In addition, Corollary \ref{corollary:f} discusses the optimal network privacy guarantee of the proposed method in various scenarios, which is a direct result of Theorem \ref{thm:consist}.

\begin{corollary}
	\label{corollary:f}
	Suppose all the conditions of Theorem 1 are met, $K = O(1)$, $d_i = \Omega(1)$ for $i \in [n]$, and $\frac{\log n}{nLs_n^2} = o(1)$.
	
	(1) If $f_i \asymp f_j$ and $f_i \gg  \big(\frac{\log n}{nLs_n^2}\big)^{1/4}$ for $i, j \in [n]$, we have $Err(\hat{\bm{c}}, \bm{c}^*) = o_p(1)$.  
	
	(2) Let $S$ denote the set of nodes such that $f_i \asymp \alpha_n$ for any $i \in S$ and $f_i \asymp 1$ otherwise, and assume $|S|/n \asymp \beta_n$. If $\frac{\beta_n}{\alpha_n^2(1-\beta_n)} = o(\frac{nLs_n^2}{\log n})$, we have $\text{Err}(\widehat{\bm{c}},\bm{c}^*) = o_p(1)$.
\end{corollary}


The first scenario of Corollary \ref{corollary:f} considers the case that all the personalized preference parameters are asymptotically of the same order. In this case, the proposed method can asymptotically reveal the network community structure as long as the personalized privacy preference parameters $f_i$ vanishes at an order slower than $\big(\frac{\log n}{nLs_n^2}\big)^{1/4}$, which further implies the differential privacy budget parameter $\epsilon_{i, j}$ should vanish at an order slower than $\sqrt{\frac{\log n}{nLs_n^2}}$ by  Lemma \ref{lemma:private_preferences}, for $1\le i \le j\le n$. The second scenario of Corollary \ref{corollary:f} considers the case where a small fraction $\beta_n$ of the nodes are highly concerned about their privacy whose privacy preference parameters are allowed to vanish at a fast order $\alpha_n$. In order to ensure the consistency of community detection, the condition $\frac{\beta_n}{\alpha_n^2(1-\beta_n)} = o(\frac{nLs_n^2}{\log n})$ is imposed to control the trade-off between $\alpha_n$ and $\beta_n$. Furthermore, the asymptotic order of the differential privacy budget parameters are categorized into three cases by Lemma \ref{lemma:private_preferences}; that is, $\epsilon_{i, j} \asymp \alpha_n^2$ if both nodes $i$ and $j$ are in $S$, $\epsilon_{i, j} \asymp \alpha_n$ if only one node $i$ or $j$ is in $S$, and $\epsilon_{i, j}\asymp 1$ if neither node $i$ nor $j$ is in $S$.


\section{Numerical experiment}
\label{Sec:Simu}
In this section, we examine the numerical performance of the proposed personalized edge-flipping mechanism in both synthetic networks and real applications.  

\subsection{Synthetic networks}

The synthetic multi-layer networks $\bm{\mathcal{A}} \in \{ 0,1 \}^{n \times n \times L}$ are generated as follows. First, the probability tensor $\bm{\mathcal{B}} \in [0,1]^{ K \times K \times L}$ is generated as $\bm{\mathcal{B}}_{k_1, k_2, l} = \big( 0.5 I(k_1=k_2) + b_{k_1, k_2, l}\big)$ with $b_{k_1, k_2, l} \sim \text{Unif}(0, 0.5)$, for $k_1, k_2 \in [K]$. Second, $\bm{c}=(c_1,\ldots,c_n)$ are randomly drawn from $[K]$ with equal probabilities, and thus obtain the resultant community assignment matrix $\bm Z$. Third, calculate $\bm{\mathcal{P}}=\bm{\mathcal{B}} \times_1 \bm{D}\bm{Z} \times_2 \bm{D}\bm{Z}$ with $d_i \sim \text{unif}(0.5,1)$ for $i \in [n]$. Finally, each entry of $\bm{\mathcal{A}}$ is generated independently according to $\bm{\mathcal{A}}_{i, j, l} \sim \text{Bernoulli}(\bm{\mathcal{P}}_{i, j, l})$, for $1 \leq i \leq j \leq n$ and $l \in [L]$.

\begin{figure}[!htb]
	\centering
	\begin{subfigure}[b]{0.45\textwidth}
		\centering
		\captionsetup{justification=centering}
		\includegraphics[width=\textwidth]{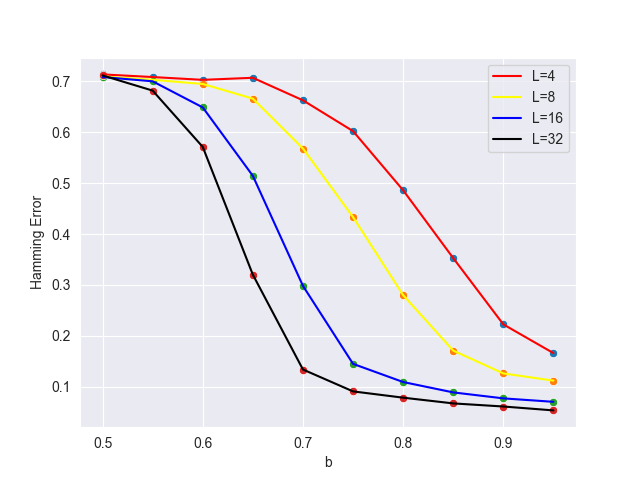}
		\caption{n=400 }
	\end{subfigure}
	\begin{subfigure}[b]{0.45\textwidth}
		\centering
		\captionsetup{justification=centering}
		\includegraphics[width=\textwidth]{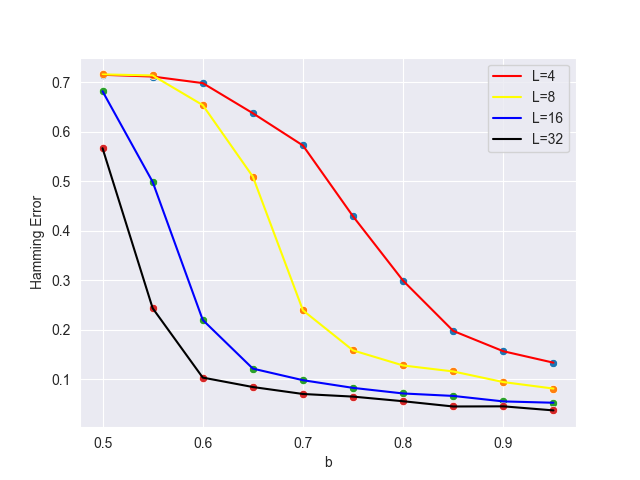}
		\caption{n=800}
	\end{subfigure}
	\captionsetup{justification=centering}
	\caption{Averaged Hamming errors over 100 replications in Example 1.}
	\label{simu1}
\end{figure}

\textbf{Example 1.} In this example, we illustrate the interplay between the accuracy of community detection and the distribution of personalized privacy parameters. To mimic the users' privacy preferences, we generate $\bm{f}$ with $f_i \sim \text{Unif}(0,b)$ and $b \in \{0.5+0.05*i: i=0, 1,...,9\}$. As for the size of multi-layer networks, we consider cases that $(n,L) \in \{400,800 \} \times \{4,8,16,32 \}$. The averaged Hamming errors over 100 replications of all cases are reported in Figure $\ref{simu1}$.

In Figure $\ref{simu1}$, as $b$ increases from 0.5 to 0.95, the Hamming errors for all values of $(n,L)$ decrease simultaneously, indicating that small personalized privacy parameters will deteriorate the community structure in multi-layer networks. In addition, when the distribution of personalized privacy parameters is fixed, the Hamming errors improve as the network size enlarges as expected.

\begin{figure}[!htb]
	\centering
	\begin{subfigure}[b]{0.45\textwidth}
		\centering
		\captionsetup{justification=centering}
		\includegraphics[width=\textwidth]{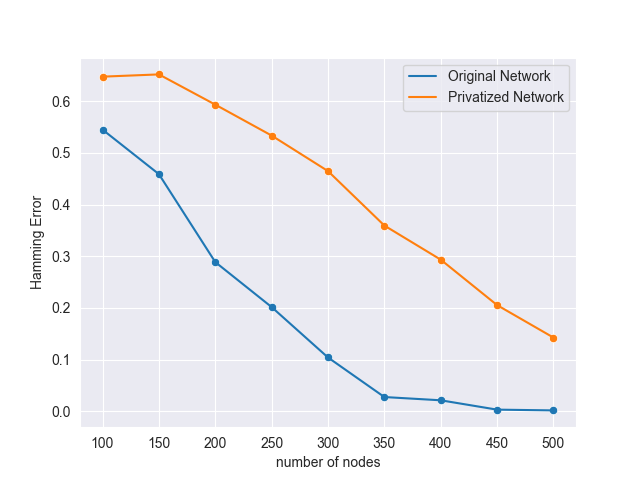}
	\end{subfigure}
	\begin{subfigure}[b]{0.45\textwidth}
		\centering
		\captionsetup{justification=centering}
		\includegraphics[width=\textwidth]{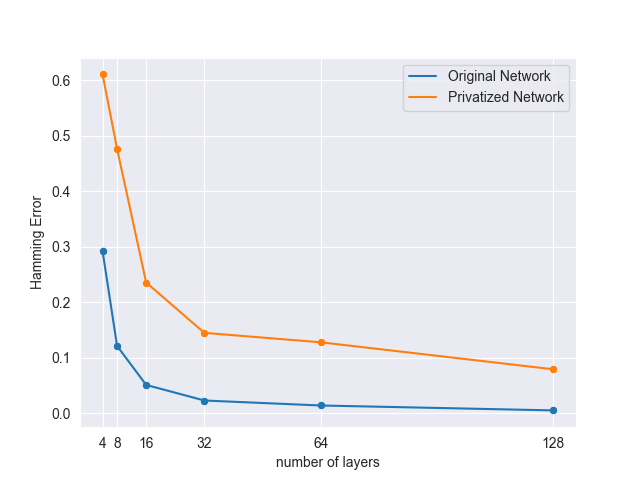}
	\end{subfigure}
	\captionsetup{justification=centering}
	\caption{	
		Averaged Hamming errors over 100 replications in Example 2.}
	\label{simu2}
\end{figure}

\textbf{Example 2.} In this example, we generate $f_i \sim \text{Unif}(0.95,1)$ for $i \in [n]$, and then consider two scenarios with increasing number of nodes or layers. Specifically, for the former scenario, we set the number of layers $L$ and the number of communities $K$ as 8 and 4, respectively, and consider cases $n \in \{100,150,200,250,...,500\}$. For the latter one, we set $(n,K)=(200,4)$ and consider $L \in  \{ 4,8,16,32,64,128\}$. The averaged Hamming errors over 100 replications of both scenarios are displayed in Figure \ref{simu2}.

Figure $\ref{simu2}$ shows that the convergence behaviors of the accuracy of community detection over privatized networks shares similar patterns as the original networks, which is consistent with the theory developed in Section 4 that community detection over privatized network maintains the similar order of convergence when personalized privacy parameters are close to 1.

\begin{figure}[!htb]
	\centering
	\includegraphics[width=\textwidth/2]{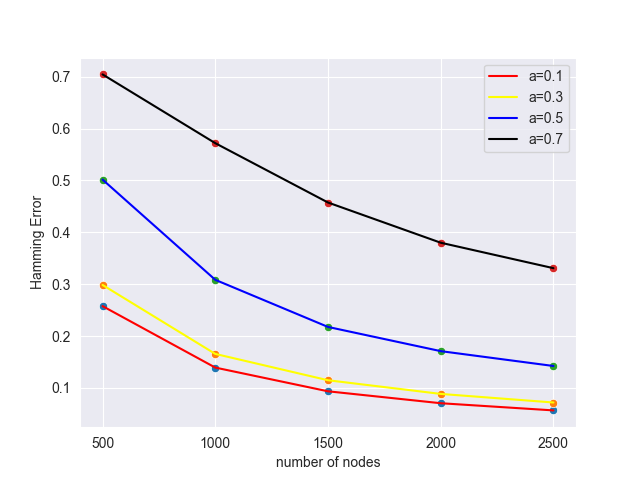}
	\caption{	
		Averaged Hamming errors over 100 replications in Example 3.}
	\label{simu3}
\end{figure}

\textbf{Example 3.} In this example, we analyze the convergence behaviors of the Hamming error when the personalized privacy parameters are polarized in that some people give up their privacy completely, whereas some users keep their connectivity behaviors as private as possible. To achieve this, we let $n^{\alpha}$ denote the number of users pursuing privacy with $a \in [0,1]$ and then we randomly sample $\lfloor 2 * n^{a} \rfloor$ nodes and set their corresponding $f_i$'s as $\sqrt{(nL)^{-1}log(n)}$ while keeping all the other $f_i$ to be 1. Moreover, we set $(K,L)=(4,4)$ and consider cases $(n,a) \in \{500,1000,1500,2000,2500 \} \times \{ 0.1,0.3,0.5,0.7 \}$. The averaged Hamming errors over 100 replications of all cases are reported in Figure \ref{simu3}.

It is evident from Figure $\ref{simu3}$ that the Hamming errors still converge when some users chose to keep their connectivity privately, and the convergence rate becomes slower when the size of these users gets larger. It suggests that, under the personalized privacy mechanism, the privacy budget can be allocated according to users' privacy preferences, and hence some users are allowed to pursue better protections of privacy in social networks.

\subsection{FriendFeed Multilayer Network}

We apply the proposed personalized edge-flipping mechanism to a FriendFeed multi-layer social network, and compare its empirical performance on the privatized network under various personalized privacy preferences. The FriendFeed network consists of a total of 574,600 interactions among 21,006 Italian users during two months' period, which is publicly available at \url{http://multilayer.it.uu.se/datasets.html}. Furthermore, the users' interactions are treated as undirected edges, and categorized into three aspects, including liking, commenting, and following, which correspond to three network layers. Since the original network layers are relatively sparse and fragile, we collect the nodes in the intersection of the giant connected components of all three network layers, and extracted the corresponding sub-graphs to create a multi-layer sub-network. This pre-processing step leads to a 3-layer network with 2,012 common nodes. 


 In social network, like the FriendFeed data, some users are not willing to reveal their friendship privacy. For example, someone might not willing to reveal her or his privacy with a famous person or a group leader in a certain community. In this case, user $i$ can choose a smaller $f_i$ to better protect her or his local connectivity pattern. Further, this can even prevent attackers from inferring user $i$'s linking pattern via transitivity. Herein, transitivity refers to the fact that a friend's friend is likely to be a friend. As such, people normally could infer the connectivity behavior between $i$ and $j$, giving their common friends $i^\prime$'s. It is thus necessary to protect the individual's local neighborhood transitivity privacy personally. Under our randomized network flipping mechanism, the users $i$'s preference is 
\begin{align*}
f_i = \sqrt{\frac{(\theta_{i^\prime, i} - \frac{1}{2})(\theta_{i, j} - \frac{1}{2})}{(\theta_{i^\prime, j} - \frac{1}{2})}}, 
\end{align*}
for any $j\ne i$, $i^\prime \ne i$ and $i^\prime \ne j$.

As $\theta_{i^\prime, i}>1/2$ by definition, $\theta_{i^\prime, i} - 1/2$ is the excess probability that $\mathcal{A}_{i, i^\prime, l}$ maintains unflipped, for $l\in [L]$. Therefore,  the larger $f_i$ is, the larger the excess maintaining probability ratio between edge pairs $(\mathcal{A}_{i^\prime, i, l}, \mathcal{A}_{i, j, l})$ and edge $\mathcal{A}_{i^\prime, j, l}$, and the transitivity pattern is more likely to maintain. If users in the FriendFeed network can choose their own preferences $f_i$'s, their local neighborhood connectivity patterns could be protected.

\begin{figure}[!htb]
	\centering
	\captionsetup{justification=centering}
	\includegraphics[width=\textwidth/2]{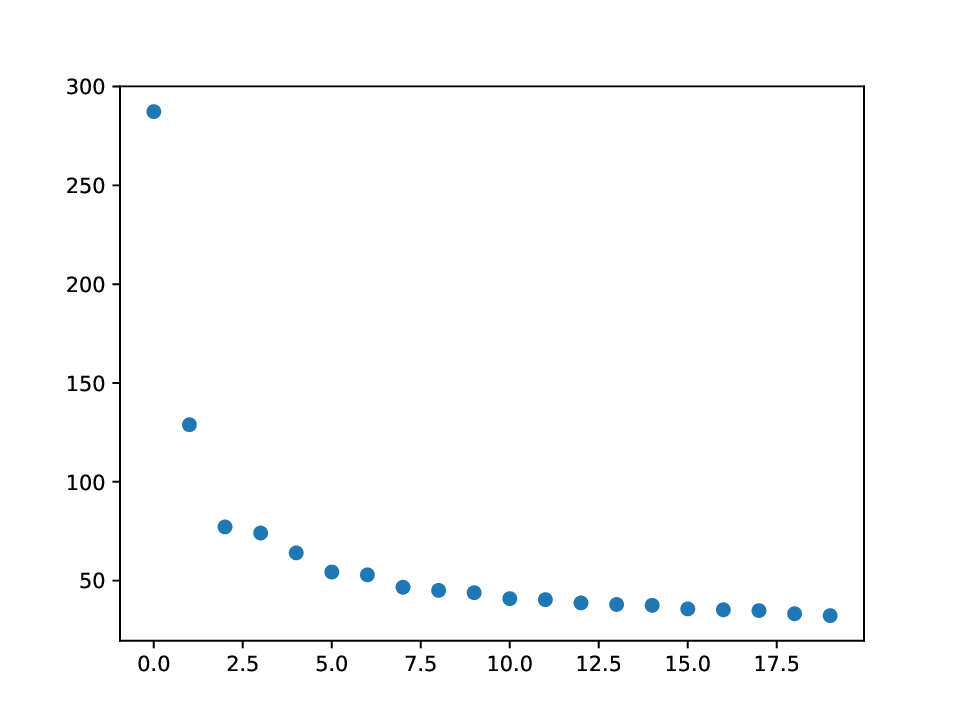}
	\captionsetup{justification=centering}
	\caption{The first 20 leading singular values of $\mathcal{M}_1(\bm{\mathcal{A}} \times_3 \bar{\bm{V}})$ in the FriendFeed multi-layer network. }
	\label{fig:svp}
\end{figure}

\begin{figure}[!htb]
	\centering
	\begin{subfigure}[b]{0.45\textwidth}
		\centering
		\captionsetup{justification=centering}
		\includegraphics[width=\textwidth]{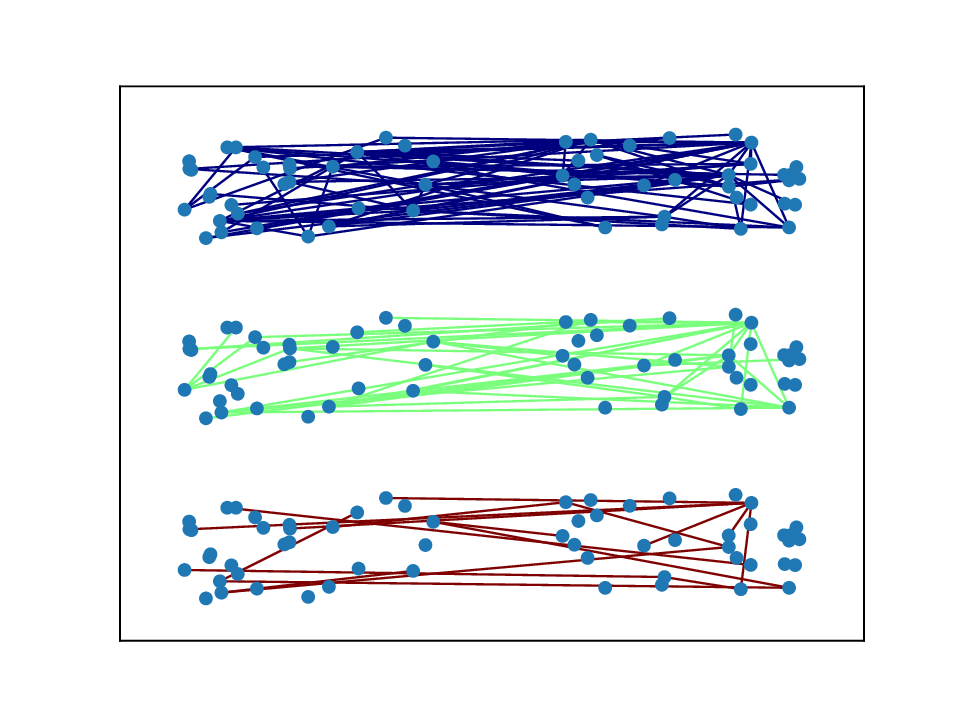}
	\end{subfigure}
	\begin{subfigure}[b]{0.45\textwidth}
		\centering
		\captionsetup{justification=centering}
		\includegraphics[width=\textwidth]{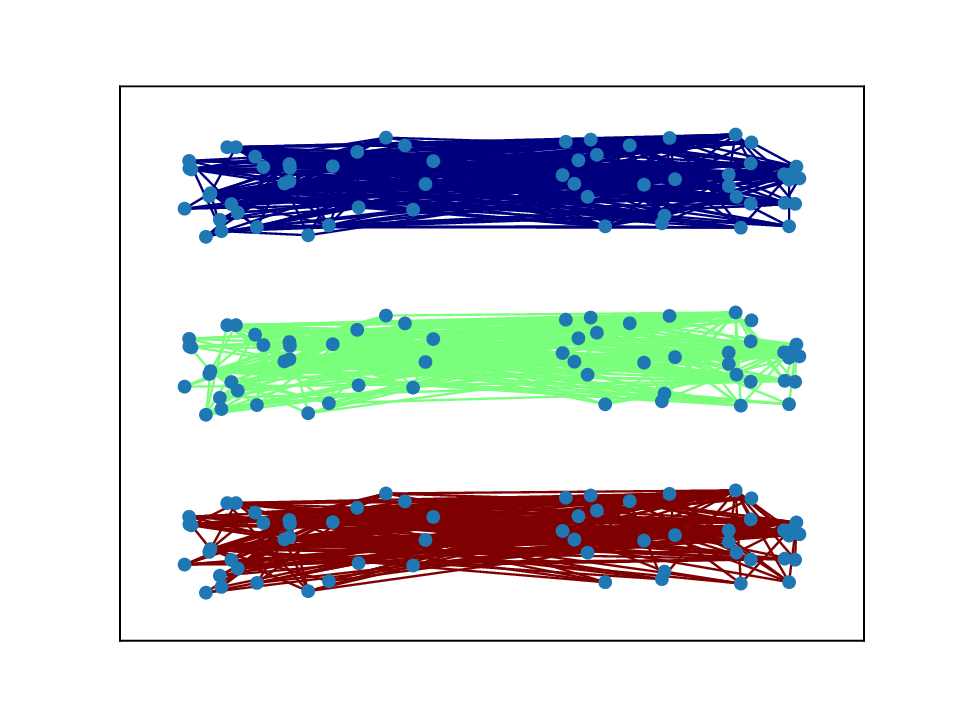}
	\end{subfigure}
	\captionsetup{justification=centering}
	\caption{The original 3-layer FriendFeed sub-network with 60 popular nodes (left), and a randomly selected flipped sub-network with $\beta = 10\%$ (right). Both panels consist of the following layer (blue), commenting layer (green), and liking layer (red).}
	\label{fig:flipped_net}
\end{figure}

Before proceeding, we first estimate the number of communities $K$ following a similar treatment as in \citet{ke2019community}. First, let $\kappa$ be a user-specific upper bound of $K$, and we perform a Tucker decomposition approximation with Tucker rank $(\kappa, \kappa, L)$ on the multi-layer network adjacency tensor $\bm{\mathcal{A}}$ to obtain mode-1 and mode-2 factor matrix $\bar{\bm{U}}$ and mode-3 factor matrix $\bar{\bm{V}}$. Next, we investigate the elbow point of the leading singular values of $\mathcal{M}_1(\bm{\mathcal{A}} \times_3 \bar{\bm{V}})$, and estimate $K$ as the number of leading singular values right before the elbow point. In the FriendFeed network, we set $\kappa=15$, and the first 20 leading singular values of $\mathcal{M}_1(\bm{\mathcal{A}} \times_3 \bar{\bm{V}})$ are displayed at Figure \ref{fig:svp}. It is clear that the elbow point appears at the 3rd leading singular value, and hence we set $K = 2$.
As there is no ground truth of the community structure in the FriendFeed netowrk, we simply treat the detected communities by the proposed method with $\bm{f}=\bm{1}_{2,012}$ as the truth. We further select $30$ nodes with the largest degrees in each detected community to visualize the 3-layer sub-network with $60$ common nodes in the left panel of Figure \ref{fig:flipped_net}. Clearly, the following layer is much denser than the other two layers, which suggests that a user may follow many other users, but only likes or comments on much fewer users she or he follows.


We then evaluate the Hamming error of the proposed method under different distributions of $\bm{f}$. To generate the personalized privacy preference vector $\bm{f}$, we randomly selected $\lfloor\beta \times 2,012\rfloor$ coordinates of $\bm{f}$ and set these privacy preference parameters as $0.02$ while setting other $f_i$'s as $0.98$, where $\lfloor x \rfloor$ denotes the largest integer that is small than or equal to $x$ and $\beta$ varies in $\{2\%, 4\%, ..., 20\%\}$. Intuitively, as $\beta$ increases, the expectation of $f_i$ decreases for $i\in [2,012]$, leading to better privacy protection for the whole network. The corresponding sub-network of a randomly selected flipped network with $\beta = 10\%$ is displayed in the right penal of Figure \ref{fig:flipped_net}. It is clear that the flipped network becomes relatively denser and substantially deviates form the original network for privacy protection. The averaged Hamming errors of the proposed method over 100 replications on the flipped FriendFeed network with various values of $\beta$ are reported in Table 1.

\begin{table}[!htbp]
	\begin{center}
		\caption{Hamming errors of the proposed method on the FriendFeed network under different edge-flipping strengths.}
		\scalebox{0.85}{
		\setlength{\tabcolsep}{4pt}
		\begin{tabular}{c|cccccccccc}
			\toprule
			$\beta$	& 2\% & 4\% & 6\% & 8\% & 10\% & 12\% & 14\% & 16\% & 18\% & 20\% \\
			\hline
			Err & 0.0723 & 0.0862 & 0.0969 & 0.1052 & 0.1235 & 0.1251 & 0.1365 & 0.1443 & 0.1501 & 0.1665 \\
			\bottomrule
		\end{tabular}
	}
	\end{center}
\end{table}
It is evident from Table 1 that the proposed method is able to deliver satisfactory community detection 
for the flipped multi-layer network the personalized edge flipping mechanism. Its Hamming errors increase with $\beta$ as expected, as the flipped networks with higher edge-flipping probabilities deviate more from the original one, leading to better privacy protection at the cost of a relatively compromised detection of communities. 

\section{Conclusions}
\label{Sec:Conc}
This paper proposes a personalized edge-flipping mechanism to protect nodes' connectivity behaviors in multi-layer network data. On the positive side, the edge flipping probabilities are allocated according to nodes' privacy preferences and demands so that protecting the connectivity behaviors could be vary from one user to the other. However, on the negative side, there might be a risk in leaking the users' privacy preferences. Theoretically, we show that the community structure of the flipped multi-layer network remains invariant under the degree-corrected multi-layer stochastic block model, which makes consistent community detection on the flipped network possible. A simple community detection method is proposed with some appropriate debiasing of the flipped network. Its asymptotic consistency is also established in terms of  community detection, which allows a small fraction of nodes to keep their connectivity behaviors as private as possible. The established theoretical results are also supported by numerical experiments on various synthetic networks and a real-life FriendFeed multi-layer network.



\par
\section*{Acknowledgements}

This work is supported in part by HK RGC Grants GRF-11301521, GRF-11311022, GRF-14306523, CUHK Startup Grant 4937091, and CUHK Direct Grant 4053588. 

\par


\bibhang=1.7pc
\bibsep=2pt
\fontsize{9}{14pt plus.8pt minus .6pt}\selectfont
\renewcommand\bibname{\large \bf References}
\expandafter\ifx\csname
natexlab\endcsname\relax\def\natexlab#1{#1}\fi
\expandafter\ifx\csname url\endcsname\relax
  \def\url#1{\texttt{#1}}\fi
\expandafter\ifx\csname urlprefix\endcsname\relax\def\urlprefix{URL}\fi

  \bibliographystyle{chicago}      
  \bibliography{ref}   

\begin{thebibliography}{}

\bibitem[\protect\citeauthoryear{Abawajy, Ninggal, and Herawan}{Abawajy
  et~al.}{2016}]{abawajy2016privacy}
Abawajy, J.~H., M.~I.~H. Ninggal, and T.~Herawan (2016).
\newblock Privacy preserving social network data publication.
\newblock {\em IEEE Communications Surveys \& Tutorials\/}~{\em 18\/}(3),
  1974--1997.

\bibitem[\protect\citeauthoryear{Alaggan, Gambs, and Kermarrec}{Alaggan
  et~al.}{2015}]{alaggan2015heterogeneous}
Alaggan, M., S.~Gambs, and A.-M. Kermarrec (2015).
\newblock Heterogeneous differential privacy.
\newblock {\em arXiv preprint arXiv:1504.06998\/}.

\bibitem[\protect\citeauthoryear{Carrington}{Carrington}{2011}]{carrington2011crime}
Carrington, P.~J. (2011).
\newblock Crime and social network analysis.
\newblock {\em The SAGE Handbook of Social Network Analysis\/}, 236--255.

\bibitem[\protect\citeauthoryear{Chen, Fung, Yu, and Desai}{Chen
  et~al.}{2014}]{chen2014correlated}
Chen, R., B.~C. Fung, P.~S. Yu, and B.~C. Desai (2014).
\newblock Correlated network data publication via differential privacy.
\newblock {\em The VLDB Journal\/}~{\em 23}, 653--676.

\bibitem[\protect\citeauthoryear{Chen, Liu, and Ma}{Chen
  et~al.}{2022}]{chen2022global}
Chen, S., S.~Liu, and Z.~Ma (2022).
\newblock Global and individualized community detection in inhomogeneous
  multilayer networks.
\newblock {\em The Annals of Statistics\/}~{\em 50\/}(5), 2664--2693.

\bibitem[\protect\citeauthoryear{Day, Li, and Lyu}{Day
  et~al.}{2016}]{day2016publishing}
Day, W.-Y., N.~Li, and M.~Lyu (2016).
\newblock Publishing graph degree distribution with node differential privacy.
\newblock In {\em Proceedings of the 2016 International Conference on
  Management of Data}, pp.\  123--138.

\bibitem[\protect\citeauthoryear{Du, Wu, Pei, Wang, and Xu}{Du
  et~al.}{2007}]{du2007community}
Du, N., B.~Wu, X.~Pei, B.~Wang, and L.~Xu (2007).
\newblock Community detection in large-scale social networks.
\newblock In {\em Proceedings of the 9th WebKDD and 1st SNA-KDD 2007 Workshop
  on Web Mining and Social Network Analysis}, pp.\  16--25.

\bibitem[\protect\citeauthoryear{Dwork, McSherry, Nissim, and Smith}{Dwork
  et~al.}{2006}]{dwork2006calibrating}
Dwork, C., F.~McSherry, K.~Nissim, and A.~Smith (2006).
\newblock Calibrating noise to sensitivity in private data analysis.
\newblock In {\em Theory of cryptography conference}, pp.\  265--284. Springer.

\bibitem[\protect\citeauthoryear{Epasto, Mirrokni, Perozzi, Tsitsulin, and
  Zhong}{Epasto et~al.}{2022}]{epasto2022differentially}
Epasto, A., V.~Mirrokni, B.~Perozzi, A.~Tsitsulin, and P.~Zhong (2022).
\newblock Differentially private graph learning via sensitivity-bounded
  personalized pagerank.
\newblock {\em Advances in Neural Information Processing Systems\/}~{\em 35},
  22617--22627.

\bibitem[\protect\citeauthoryear{Fan, Zhang, and Yan}{Fan
  et~al.}{2020}]{fan2020asymptotic}
Fan, Y., H.~Zhang, and T.~Yan (2020).
\newblock Asymptotic theory for differentially private generalized
  $\beta$-models with parameters increasing.
\newblock {\em arXiv preprint arXiv:2002.12733\/}.

\bibitem[\protect\citeauthoryear{Ghoshdastidar and Dukkipati}{Ghoshdastidar and
  Dukkipati}{2017}]{ghoshdastidar2017uniform}
Ghoshdastidar, D. and A.~Dukkipati (2017).
\newblock Uniform hypergraph partitioning: Provable tensor methods and sampling
  techniques.
\newblock {\em The Journal of Machine Learning Research\/}~{\em 18\/}(1),
  1638--1678.

\bibitem[\protect\citeauthoryear{Granovetter}{Granovetter}{2005}]{granovetter2005impact}
Granovetter, M. (2005).
\newblock The impact of social structure on economic outcomes.
\newblock {\em Journal of Economic Perspectives\/}~{\em 19\/}(1), 33--50.

\bibitem[\protect\citeauthoryear{Gregurec, Vrane{\v{s}}evi{\'c}, and
  Dobrini{\'c}}{Gregurec et~al.}{2011}]{gregurec2011importance}
Gregurec, I., T.~Vrane{\v{s}}evi{\'c}, and D.~Dobrini{\'c} (2011).
\newblock The importance of database marketing in social network advertising.
\newblock {\em International Journal of Management Cases\/}~{\em 13\/}(4),
  165--172.

\bibitem[\protect\citeauthoryear{Guo, Qiu, Zhang, and Chang}{Guo
  et~al.}{2020}]{guo2020randomized}
Guo, X., Y.~Qiu, H.~Zhang, and X.~Chang (2020).
\newblock Randomized spectral co-clustering for large-scale directed networks.
\newblock {\em arXiv preprint arXiv:2004.12164\/}.

\bibitem[\protect\citeauthoryear{Hay, Li, Miklau, and Jensen}{Hay
  et~al.}{2009}]{hay2009accurate}
Hay, M., C.~Li, G.~Miklau, and D.~Jensen (2009).
\newblock Accurate estimation of the degree distribution of private networks.
\newblock In {\em 2009 Ninth IEEE International Conference on Data Mining},
  pp.\  169--178. IEEE.

\bibitem[\protect\citeauthoryear{Hehir, Slavkovi{\'c}, and Niu}{Hehir
  et~al.}{2022}]{hehir2022consistent}
Hehir, J., A.~Slavkovi{\'c}, and X.~Niu (2022).
\newblock Consistent spectral clustering of network block models under local
  differential privacy.
\newblock {\em The Journal of privacy and confidentiality\/}~{\em 12\/}(2).

\bibitem[\protect\citeauthoryear{Ji, Li, Srivatsa, He, and Beyah}{Ji
  et~al.}{2014}]{ji2014structure}
Ji, S., W.~Li, M.~Srivatsa, J.~S. He, and R.~Beyah (2014).
\newblock Structure based data de-anonymization of social networks and mobility
  traces.
\newblock In {\em International Conference on Information Security}, pp.\
  237--254. Springer.

\bibitem[\protect\citeauthoryear{Jin}{Jin}{2015}]{jin2015fast}
Jin, J. (2015).
\newblock Fast community detection by score.
\newblock {\em The Annals of Statistics\/}~{\em 43\/}(1), 57--89.

\bibitem[\protect\citeauthoryear{Jing, Li, Lyu, and Xia}{Jing
  et~al.}{2021}]{jing2020community}
Jing, B.-Y., T.~Li, Z.~Lyu, and D.~Xia (2021).
\newblock Community detection on mixture multilayer networks via regularized
  tensor decomposition.
\newblock {\em The Annals of Statistics\/}~{\em 49\/}(6), 3181--3205.

\bibitem[\protect\citeauthoryear{Karwa and Slavkovi{\'c}}{Karwa and
  Slavkovi{\'c}}{2016}]{karwa2016inference}
Karwa, V. and A.~Slavkovi{\'c} (2016).
\newblock Inference using noisy degrees: Differentially private $\beta$-model
  and synthetic graphs.
\newblock {\em The Annals of Statistics\/}~{\em 44\/}(1), 87--112.

\bibitem[\protect\citeauthoryear{Kasiviswanathan, Nissim, Raskhodnikova, and
  Smith}{Kasiviswanathan et~al.}{2013}]{kasiviswanathan2013analyzing}
Kasiviswanathan, S.~P., K.~Nissim, S.~Raskhodnikova, and A.~Smith (2013).
\newblock Analyzing graphs with node differential privacy.
\newblock In {\em Theory of Cryptography Conference}, pp.\  457--476. Springer.

\bibitem[\protect\citeauthoryear{Ke, Shi, and Xia}{Ke
  et~al.}{2019}]{ke2019community}
Ke, Z.~T., F.~Shi, and D.~Xia (2019).
\newblock Community detection for hypergraph networks via regularized tensor
  power iteration.
\newblock {\em arXiv preprint arXiv:1909.06503\/}.

\bibitem[\protect\citeauthoryear{Klerks}{Klerks}{2004}]{klerks2004network}
Klerks, N.~P. (2004).
\newblock The network paradigm applied to criminal organisations: theoretical
  nitpicking or a relevant doctrine for investigators? recent developments in
  the: Theoretical nitpicking or a relevant doctrine for investigators? recent.
\newblock In {\em Transnational Organised Crime}, pp.\  111--127. Routledge.

\bibitem[\protect\citeauthoryear{Kolda and Bader}{Kolda and
  Bader}{2009}]{kolda2009tensor}
Kolda, T.~G. and B.~W. Bader (2009).
\newblock Tensor decompositions and applications.
\newblock {\em SIAM review\/}~{\em 51\/}(3), 455--500.

\bibitem[\protect\citeauthoryear{Lei, Chen, and Lynch}{Lei
  et~al.}{2020}]{lei2020consistent}
Lei, J., K.~Chen, and B.~Lynch (2020).
\newblock Consistent community detection in multi-layer network data.
\newblock {\em Biometrika\/}~{\em 107\/}(1), 61--73.

\bibitem[\protect\citeauthoryear{Lei and Rinaldo}{Lei and
  Rinaldo}{2015}]{lei2015consistency}
Lei, J. and A.~Rinaldo (2015).
\newblock Consistency of spectral clustering in stochastic block models.
\newblock {\em The Annals of Statistics\/}~{\em 43\/}(1), 215--237.

\bibitem[\protect\citeauthoryear{Leskovec, Lang, and Mahoney}{Leskovec
  et~al.}{2010}]{leskovec2010empirical}
Leskovec, J., K.~J. Lang, and M.~Mahoney (2010).
\newblock Empirical comparison of algorithms for network community detection.
\newblock In {\em Proceedings of the 19th International Conference on World
  Wide Web}, pp.\  631--640.

\bibitem[\protect\citeauthoryear{Li and Das}{Li and
  Das}{2013}]{li2013applications}
Li, N. and S.~K. Das (2013).
\newblock Applications of k-anonymity and $l$-diversity in publishing online
  social networks.
\newblock In {\em Security and Privacy in Social Networks}, pp.\  153--179.
  Springer.

\bibitem[\protect\citeauthoryear{Ma and Nandy}{Ma and
  Nandy}{2023}]{ma2023community}
Ma, Z. and S.~Nandy (2023).
\newblock Community detection with contextual multilayer networks.
\newblock {\em IEEE Transactions on Information Theory\/}~{\em 69\/}(5),
  3203--3239.

\bibitem[\protect\citeauthoryear{Mohamed, Nguyen, Vullikanti, and
  Tandon}{Mohamed et~al.}{2022}]{mohamed2022differentially}
Mohamed, M.~S., D.~Nguyen, A.~Vullikanti, and R.~Tandon (2022).
\newblock Differentially private community detection for stochastic block
  models.
\newblock In {\em International Conference on Machine Learning}, pp.\
  15858--15894. PMLR.

\bibitem[\protect\citeauthoryear{Nayak and Adeshiyan}{Nayak and
  Adeshiyan}{2009}]{nayak2009unified}
Nayak, T.~K. and S.~A. Adeshiyan (2009).
\newblock A unified framework for analysis and comparison of randomized
  response surveys of binary characteristics.
\newblock {\em Journal of Statistical Planning and Inference\/}~{\em 139\/}(8),
  2757--2766.

\bibitem[\protect\citeauthoryear{Paul and Chen}{Paul and
  Chen}{2021}]{paul2021null}
Paul, S. and Y.~Chen (2021).
\newblock Null models and community detection in multi-layer networks.
\newblock {\em Sankhya A\/}, 1--55.

\bibitem[\protect\citeauthoryear{Rudelson and Vershynin}{Rudelson and
  Vershynin}{2009}]{rudelson2009smallest}
Rudelson, M. and R.~Vershynin (2009).
\newblock Smallest singular value of a random rectangular matrix.
\newblock {\em Communications on Pure and Applied Mathematics: A Journal Issued
  by the Courant Institute of Mathematical Sciences\/}~{\em 62\/}(12),
  1707--1739.

\bibitem[\protect\citeauthoryear{Thomas and Nicol}{Thomas and
  Nicol}{2010}]{thomas2010koobface}
Thomas, K. and D.~M. Nicol (2010).
\newblock The koobface botnet and the rise of social malware.
\newblock In {\em 2010 5th International Conference on Malicious and Unwanted
  Software}, pp.\  63--70. IEEE.

\bibitem[\protect\citeauthoryear{Ullman and Sealfon}{Ullman and
  Sealfon}{2019}]{ullman2019efficiently}
Ullman, J. and A.~Sealfon (2019).
\newblock Efficiently estimating erdos-renyi graphs with node differential
  privacy.
\newblock {\em Advances in Neural Information Processing Systems\/}~{\em 32},
  3770--3780.

\bibitem[\protect\citeauthoryear{Wang, Yan, Jiang, and Leng}{Wang
  et~al.}{2022}]{wang2022two}
Wang, Q., T.~Yan, B.~Jiang, and C.~Leng (2022).
\newblock Two-mode networks: inference with as many parameters as actors and
  differential privacy.
\newblock {\em Journal of Machine Learning Research\/}~{\em 23\/}(292), 1--38.

\bibitem[\protect\citeauthoryear{Wang, Wu, and Hu}{Wang
  et~al.}{2016}]{wang2016using}
Wang, Y., X.~Wu, and D.~Hu (2016).
\newblock Using randomized response for differential privacy preserving data
  collection.
\newblock In {\em EDBT/ICDT Workshops}, Volume 1558, pp.\  0090--6778.

\bibitem[\protect\citeauthoryear{Xu, Yuan, Wu, and Phan}{Xu
  et~al.}{2018}]{xu2018dpne}
Xu, D., S.~Yuan, X.~Wu, and H.~Phan (2018).
\newblock Dpne: Differentially private network embedding.
\newblock In {\em Advances in Knowledge Discovery and Data Mining: 22nd
  Pacific-Asia Conference, PAKDD 2018, Melbourne, VIC, Australia, June 3-6,
  2018, Proceedings, Part II 22}, pp.\  235--246. Springer.

\bibitem[\protect\citeauthoryear{Xu, Zhen, and Wang}{Xu
  et~al.}{2023}]{xu2023covariate}
Xu, S., Y.~Zhen, and J.~Wang (2023).
\newblock Covariate-assisted community detection in multi-layer networks.
\newblock {\em Journal of Business \& Economic Statistics\/}~{\em 41\/}(3),
  915--926.

\bibitem[\protect\citeauthoryear{Yan}{Yan}{2021}]{yan2021directed}
Yan, T. (2021).
\newblock Directed networks with a differentially private bi-degree sequence.
\newblock {\em Statistica Sinica\/}~{\em 31\/}(4), 2031--2050.

\bibitem[\protect\citeauthoryear{Yan}{Yan}{2023}]{yan2023differentially}
Yan, T. (2023).
\newblock Differentially private analysis of networks with covariates via a
  generalized $\beta$-model.
\newblock {\em arXiv preprint arXiv:2311.10279\/}.

\bibitem[\protect\citeauthoryear{Zhen and Wang}{Zhen and
  Wang}{2023}]{zhen2021community}
Zhen, Y. and J.~Wang (2023).
\newblock Community detection in general hypergraph via graph embedding.
\newblock {\em Journal of the American Statistical Association\/}~{\em
  118\/}(543), 1620--1629.

\end{thebibliography}

%
%
%
%
%

\vskip .65cm
\noindent
Department of Statistics, The Chinese University of Hong Kong
\vskip 2pt
\noindent
E-mail: yzhen8-c@my.cityu.edu.hk
\vskip 2pt

\noindent
 Department of Statistics and Data Science, University of California, Los Angeles
\vskip 2pt
\noindent
E-mail: shirong@stat.ucla.edu

\noindent
Department of Statistics, The Chinese University of Hong Kong
\vskip 2pt
\noindent
E-mail: junhuiwang@cuhk.edu.hk

\end{document}